\documentclass[submission,copyright,creativecommons]{eptcs}
\providecommand{\event}{FMAS 2022} 

\usepackage{iftex}
\usepackage{graphicx}
\usepackage[colorinlistoftodos,prependcaption,textsize=tiny]{todonotes}
\usepackage{tikz}
\usetikzlibrary{shapes.geometric, arrows}
\tikzstyle{startstop} = [rectangle, rounded corners, minimum width=4cm, minimum height=0.6 cm,text centered, draw=black, fill=red!30]
\tikzstyle{io} = [trapezium, trapezium left angle=70, trapezium right angle=110, minimum width=3cm, minimum height=0.6 cm, text centered, draw=black, fill=blue!30]
\tikzstyle{process} = [rectangle, minimum width=3cm, minimum height=0.6 cm, text centered, draw=black, fill=orange!30]
\tikzstyle{process2} = [rectangle, minimum width=3cm, minimum height=0.6 cm, text centered, draw=black, fill=red!30]
\tikzstyle{process3} = [rectangle, minimum width=1.5cm, minimum height=0.6 cm, text centered, draw=black, fill=green!30]
\tikzstyle{process4} = [rectangle, minimum width=1.5cm, minimum height=0.6 cm, text centered, draw=black, fill=blue!30]
\tikzstyle{process5} = [rectangle, minimum width=1.5cm, minimum height=0.6 cm, text centered, draw=black, fill=yellow!30]
\tikzstyle{process01} = [rectangle, minimum width=1.5cm, minimum height=0.6 cm, text centered, draw=black, fill=orange!30]
\tikzstyle{process21} = [rectangle, minimum width=1.5cm, minimum height=0.6 cm, text centered, draw=black, fill=red!30]
\tikzstyle{process31} = [rectangle, minimum width=1.5cm, minimum height=0.6 cm, text centered, draw=black, fill=green!30]
\tikzstyle{process41} = [rectangle, minimum width=1.5cm, minimum height=0.6 cm, text centered, draw=black, fill=blue!30]
\tikzstyle{process51} = [rectangle, minimum width=1.5cm, minimum height=0.6 cm, text centered, draw=black, fill=yellow!30]
\tikzstyle{decision} = [diamond, minimum width=2cm, minimum height=0.6 cm, text width=3cm, text centered, draw=black, fill=green!30]
\tikzstyle{block} = [rectangle, minimum width=0.6cm, minimum height=0.6cm, text centered, draw=black, fill=blue!30]
\tikzstyle{block2} = [rectangle, minimum width=0.6cm, minimum height=0.6cm, text centered, draw=black, fill=red!30]
\tikzstyle{block3} = [rectangle, minimum width=0.6cm, minimum height=0.6cm, text centered, draw=black, fill=green!30]
\tikzstyle{block4} = [rectangle, minimum width=0.6cm, minimum height=0.6 cm, text centered, draw=black, fill=yellow!30]
\tikzstyle{block5} = [rectangle, minimum width=0.6 cm, minimum height=0.6 cm, text centered, draw=black, fill=orange!30]
\tikzstyle{block6} = [rectangle, minimum width=0.6 cm, minimum height=0.6 cm, text centered, draw=black, fill=pink!30]
\tikzstyle{block7} = [rectangle, minimum width=0.6 cm, minimum height=0.6 cm, text centered, draw=black, fill=white!30]
\tikzstyle{arrow} = [thick,->,>=stealth]
\usepackage{adjustbox}
\usepackage{listings}
\ifpdf
  \usepackage{underscore}         
  \usepackage[T1]{fontenc}        
\else
  \usepackage{breakurl}           
\fi

\title{Generating Safe Autonomous Decision-Making in ROS}
\author{
Yi Yang \qquad\qquad Tom Holvoet
\institute{imec-DistriNet\\
Dept. of Computer Science\\
KU Leuven\\
B-3001 Leuven, Belgium}
\email{\quad yi.yang@kuleuven.be \quad\qquad tom.holvoet@kuleuven.be}
}
\def\titlerunning{Generating Safe Autonomous Decision-Making in ROS}
\def\authorrunning{Yi Yang \& Tom Holvoet}
\begin{document}
\maketitle

\begin{abstract}
The Robot Operating System (ROS) is a widely used framework for building robotic systems. It offers a wide variety of reusable packages and a pattern for new developments. It is up to developers how to combine these elements and integrate them with decision-making for autonomous behavior. The feature of such decision-making that is in general valued the most is safety assurance.

In this research preview, we present a formal approach for generating safe autonomous decision-making in ROS. We first describe how to improve our existing static verification approach to verify multi-goal multi-agent decision-making. After that, we describe how to transition from the improved static verification approach to the proposed runtime verification approach. An initial implementation of this research proposal yields promising results.
\end{abstract}

\section{Introduction}
Autonomous systems have the ability to make decisions to achieve predefined goals without human or other guidance or control. As such, the heart of autonomous systems is their autonomous decision-making. Throughout this article, we use an autonomous mobile robot as an example. This autonomous mobile robot is set to have specific goals, some prior knowledge of the world, and safety specifications. Ideally, it should autonomously make decisions that allow the system to achieve its goals while fully adhering to all provided safety specifications.
The most important requirements for autonomous decision-making are effective behaviors and safety assurance.

The Robot Operating System (ROS) \cite{quigley2009ros} is a collection of software libraries and tools for developing robot applications. Due to its distinctive features—modularity, extensive functional libraries, and open source, ROS has become the most well-known framework for constructing robotic systems. In ROS, a number of frequently used functions, including navigation, manipulation, and locomotion, have been carefully developed as reusable packages. These functions can be deployed by engineers, thereby reducing the burden of development. Notably, the most well-developed functions are the implementation of robotic actions, yet there is still significant space for advancement in the decision-making of robots. 

The safety of autonomous systems mainly refers to safe decision-making in autonomous systems. For decision-making, we are studying agent programming languages \cite{shoham1993agent}. Compared with other agent programming languages, GOAL \cite{hindriks2009programming} is particularly well suited for programming autonomous decision-making, because it generates decisions during execution rather than relying on predefined plan specifications. Formal verification is a convincing approach for providing safety assurance of high-confidence systems \cite{woodcock2009formal}. Formal verification can be categorized into static verification and runtime verification. Static verification verifies the given properties of all possible runs of a program, while runtime verification monitors one execution of a program and verifies properties at runtime \cite{ahrendt2012unified}. 

This is a research preview that outlines a well-defined research idea of generating safe autonomous decision-making for autonomous systems developed using ROS, combining ROS, decision-making, and formal verification. The research idea is based on our static verification approach for verifying autonomous decision-making. We have two contributions: first, we explain how to improve the existing static verification approach to verify multi-goal multi-agent decision-making; second, we describe how to transition from the static verification approach to the runtime verification approach. The paper is structured as follows. In Section 2, we introduce the related work to our work. In Section 3, we briefly explain how to improve our static verification approach to verify multi-goal multi-agent autonomous decision-making. In Section 4, we introduce the rationale, the initial design of the architecture of a decision-making node, and the analysis of the proposed runtime verification approach for generating safe autonomous decision-making in ROS. In Section 5, we draw conclusions and indicate future work. 

\section{Related Work}
While model checking is popular in verifying agent programming languages \cite{bordini2003model} \cite{dennis2012model}, model checking is not a prevalent approach in verifying GOAL. GOAL has many features in common with BDI APLs \cite{de2020bdi}, but it focuses on logical derivation, making it ideal for deductive verification. The inventors of GOAL proposed the first verification framework for GOAL in \cite{de2007verification}. This paper demonstrates how to conduct verification for GOAL programs.  \cite{jensen2021using} implemented the verification framework proposed in \cite{de2007verification} using Isabelle \cite{nipkow2002isabelle}. Isabelle's implementation provides high confidence for the verification framework, but has limitations: first, it can only specify a subset of GOAL programs, with only a single agent, a single goal, propositional features, and no communication; second, it lacks an automated verification process, so users have to be familiar with Isabelle. 

As no well-developed ROS package for generating decision-making is widely accepted by most users, decision-making remains a challenging issue in ROS. In general, there are two approaches to generating decision-making in ROS: providing a decision-making node in ROS, and connecting an agent programming language to ROS.

The conventional approach to generating decision-making in ROS is to provide a decision-making node. Both packages, i.e. \cite{DM2019} and \cite{MDM2015}, support the generation of decisions in ROS. Both packages do not use formal verification to generate decisions. In other words, neither of these two packages can ensure that ROS will generate safe decisions. Moreover, the decision-making generation procedures in both packages are problem-dependent, i.e., each unique problem necessitates a unique implementation of the decision-making generation process, which adds to the development difficulties for users.

For the purpose of programming complex decision-making, agent programming languages have been developed. Many efforts have been made to provide safety assurance for the generated decisions by agent programming languages \cite{de2007verification} \cite{bordini2003model} \cite{dennis2012model} \cite{holzmann1997model}. By integrating an agent programming language with ROS, the decision-making in ROS can benefit from the existing research findings on agent programming languages.

\cite{cardoso2020interface} presents an interface for the Gwendolen agent programming language \cite{dennis2008gwendolen} to communicate with ROS. Moreover, the AJPF model checker \cite{dennis2012model} is used to verify the properties of the decisions generated by the Gwendolen language. Therefore, the interface makes the generation of verified decisions feasible in ROS. However, no performance comparison has been conducted between the presented approach in \cite{cardoso2020interface} and the traditional approach. Rosbridge \cite{crick2017rosbridge} is the primary tool used to implement the interface, but Rosbridge is particularly designed for facilitating web applications with ROS \cite{crick2017rosbridge}. The connection of an agent programming language to ROS is not quite in accordance with the design intent of Rosbridge, which may result in certain performance concerns. \cite{martin2018generic} describes a generic multi-layer architecture based on ROS-JADE integration for autonomous transport vehicles. The cognitive layer of the proposed multi-layer architecture is essentially a ROS-JADE node, serving as a communicating function between ROS and JADE \cite{bellifemine2005jade}. This paper focuses on the generic multi-layer architecture instead of the details of the cognitive layer. The validation of real settings and the efficiency issues remain for their future work. Considering the architectural design of the ROS-JADE node, it lacks safety assurance.

As a lightweight verification approach, runtime verification is usually considered ideal to provide safety assurance for complicated systems, such as robotic applications. Many runtime verification frameworks have been proposed, such as ROSRV \cite{huang2014rosrv} and ROSMonitoring \cite{ferrando2020rosmonitoring}. ROSRV and ROSMonitoring provide safety assurance for autonomous systems by intercepting those messages that violate safety requirements.

\section{Static Verification}
Our prior work \cite{yi2022} serves as the foundation for the proposed research idea. GOAL sets itself apart from the majority of agent programming languages due to the automated decision-making process. This distinctive feature was the primary factor in our decision to adopt it as the instrument to program autonomous decision-making. \cite{yi2022} is the initial proposal for verifying autonomous decision-making by imposing single-goal and single-agent restrictions on the GOAL program. \cite{yi2022} presents a justified algorithm transforming a stratified GOAL program for a single agent and a single goal to its equivalent transition system, equivalent in terms of operational semantics.
We now briefly describe how to substantially advanced the work presented in \cite{yi2022} by removing these constraints on GOAL programs. The proposed static verification approach is presumably to verify multi-goal multi-agent decision-making in GOAL. 

\begin{figure}[h]
   \centering
    \begin{tikzpicture}[node distance=1.2cm]
\node (1) [process01] {FO Logic};
\node (2) [process21, right of=1,yshift=+2.4cm,xshift=+2cm] {Property};
\node (3) [process21, below of=2 ] {Constraint};
\node (4) [process21, below of=3, xshift=-1cm ] {Sent Msg};
\node (5) [process21, below of=3, xshift=+1cm ] {Action};
\node (6) [process21, below of=4, xshift=+1cm ] {Received Msg};
\node (7) [process21, below of=6 ] {Substate Update};
\node (8) [process31, left of=1, yshift=+2cm, xshift=-2cm ] {Specification};
\node (9) [process41, below of=8 ] {Encoding};
\node (10) [process51, below of=9, yshift=-0.5cm ] {Prism Model};
\draw [arrow] (1) to [out=90,in=180] (2.west);
\draw [arrow] (1) to [out=45,in=180] (3.west);
\draw [arrow] (1) to [out=0,in=180] (4.west);
\draw [arrow] (1) to [out=20,in=150] (5);
\draw [arrow] (1) to [out=300,in=180] (6.west);
\draw [arrow] (2) -- (3);
\draw [arrow] (3) -- (4);
\draw [arrow] (3) -- (5);
\draw [arrow] (4) to [out=220,in=180] (7.west);
\draw [arrow] (5) to [out=330,in=0] (7.east);
\draw [arrow] (6) -- (7);
\draw (-1,-3) rectangle (5.5,3);
\draw [arrow] (8) -- (-1,2);
\draw [arrow] (-1,0.8) -- (9);
\draw (0.2,2.8) node{TS Generation};
\draw [arrow] (9) -- (10);
\end{tikzpicture}
    \caption{Automated translator for GOAL (Architecture)}
    \label{fig:ATG}
\end{figure}
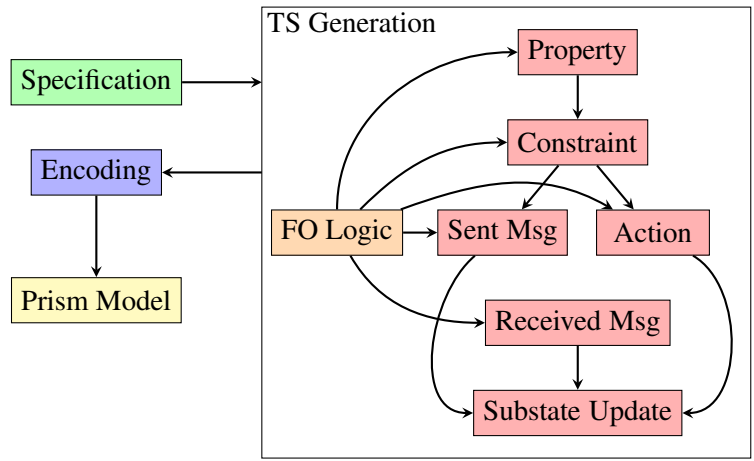

Figure \ref{fig:ATG} shows the architecture of the automated translator. The automated translator consists of two main modules: transition system (TS) Generation and Encoding. The input of the automated translator is the specifications of a multi-agent system that can be specified with GOAL. The output of the automated translator is a Prism model. The automated translator transforms the specifications into its operational-semantically equivalent Prism model that can be directly processed by two efficient probabilistic symbolic model checkers: Storm \cite{dehnert2017storm} and Prism \cite{kwiatkowska2002prism}. 

The specification language is an extended version of the specification language provided in \cite{yi2022}. The specification language is inherited from the the formalization of the syntax and semantics of the internal logic of GOAL. The first-order (FO) logic used for the formalization can properly represent the essential characteristics of GOAL programs, including durative actions, multiple goals, multiple agents, and agent communication. The FO logic employed for the formalization serves as a specification language. The specification language can properly specify a multi-agent system in a modular way. We point out that the specification language is suitable for expressing the decision-making generating processes of autonomous systems in addition to GOAL programs. In other words, the specification language can be used to specify the decision-making generating process rather than GOAL. The specifications can be categorized as system-level specifications and agent-level specifications. System-level specifications include knowledge base, communication rules, event processing rules, action generation rules, domains for multi-variables, and constants. Agent-level specifications describe the belief bases and the goal base of an agent. 

The crucial step of verifying GOAL programs is the transformation from the GOAL programs to an operational-semantically equivalent transition system. We briefly explain the key concepts of the transformation. In GOAL, the mental state of an agent is a pair consisting of its beliefs and goals \cite{de2007verification}. For a GOAL program for a multi-agent system with n agents, each agent has its own mental state. The state of the GOAL program is an n-tuple, whose element is a substate. Precisely, the mental state of an agent is a substate of the GOAL program for a multi-agent system. Each mental state is associated with substate properties. This property is the minimal model of the FO theory consisting of the beliefs and the knowledge base. Based on the logical derivation and the minimal model generation, feasible decisions are generated. A generated decision of a GOAL program is mapped to an action of the transition system. Executed decisions can modify the mental state of an agent. We can conclude that a GOAL program only generates safe decision-making if the generated operational-semantically equivalent transition system satisfies all given safety properties. This is predicated on the assumption that the specification of the GOAL program properly renders the internal logic of the GOAL program. Notably, we use the automated decision-making mechanism to implement the transformation, and the key components of the automated decision-making mechanism are automated logical derivation and minimal model generation. Instead of being problem-dependent, the decision-making generation is logic-dependent, which makes it easier for users to conduct verification.

\section{Runtime Verification Approach}
\subsection{Rationale}
The improved static verification approach can provide autonomous systems with the result of PCTL model checking analysis. Assuming that the specifications accurately represent the internal logic of the autonomous decision-making, the static verification approach offers advanced verification results, including convincing safety assurance for the decision-making generation process. However, the static verification approach suffers from certain inherent limitations, such as processing real-time information and the state-space explosion. Based on the existing static verification approach, we propose a runtime verification approach to complement the existing static verification approach. The main component of the proposed runtime verification approach is a decision-making node in ROS. We draw attention to the fact that the runtime verification of autonomous decision-making can benefit from the specification language of the automated translator and the automated decision-making generation.

\begin{table}[h]
    \centering
    \begin{tabular}{ |c|c|c| }
 \hline
 Comparison& Static Verification& Runtime Verification\\
 \hline
  \hline
 DM Generation&Same&Same\\
 \hline
 Specification language&Same&Same\\
  \hline
  Specification Provider&User&User and Sensor\\
  \hline
  State Space&All States &One State\\
 \hline
  Analysis&PCTL model checking&Safety Checking\\
 \hline
\end{tabular}
    \caption{Comparison}
    \label{tab:1}
\end{table}
In summary, Table \ref{tab:1} presents a comparison between the improved static verification approach and the proposed runtime verification approach. Both approaches share the same implementation of the decision-making generation, and they use the same specification language as the input. Users should provide all specifications in the static verification approach, whereas users and sensors provide all specifications in the runtime verification approach. While the runtime verification approach only generates one state at a time, the static verification approach always generates the entire state space. While the static verification approach can perform PCTL model checking, the proposed runtime verification approach only checks the safety at runtime. As the existing static verification approach cannot handle real-time information, this will raise safety concerns at runtime. As a complementary approach, the proposed runtime verification approach focuses on safety checking.

\subsection{Architecture for a Decision-Making Node}
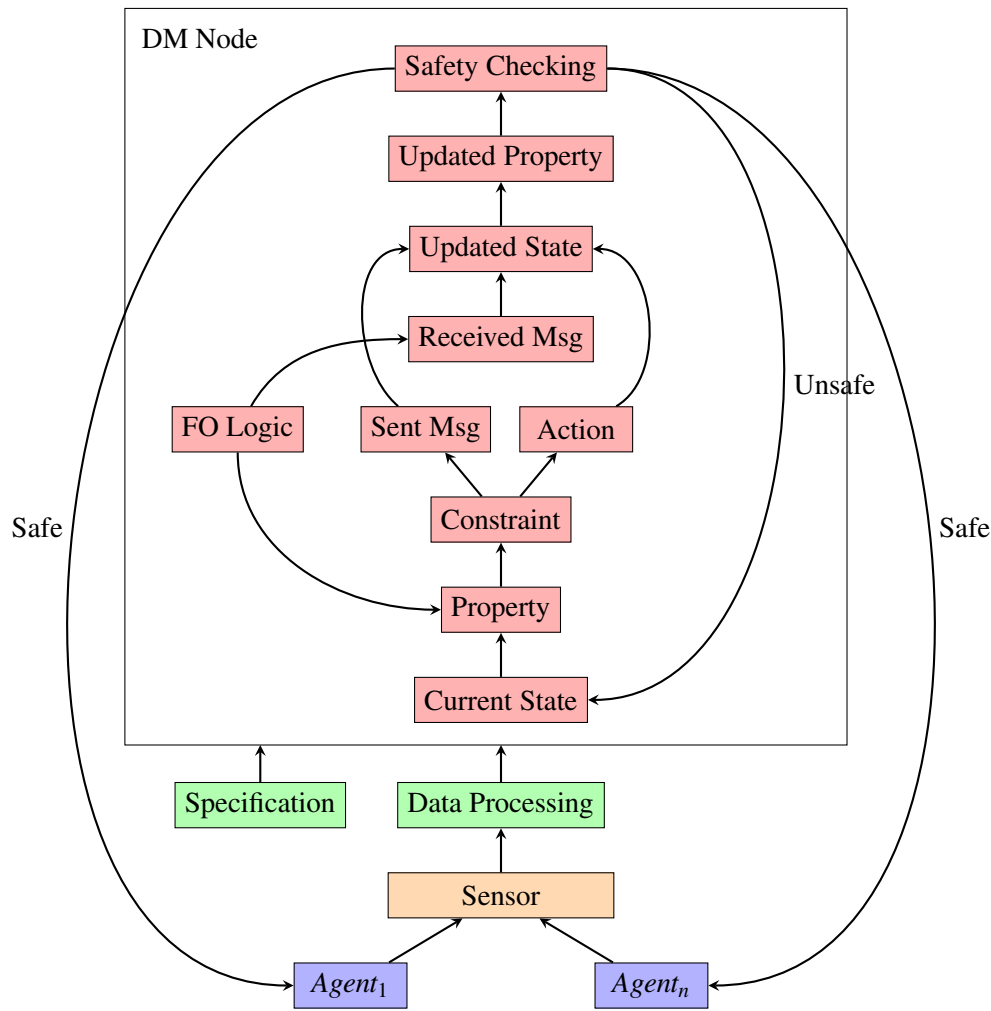
\begin{figure}[h]
   \centering
   \vspace{-2mm}
    \begin{tikzpicture}[node distance=1.2cm]
\node (1) [process3] {Data Processing};
\node (3) [process3, left of =1,xshift=-2cm] {Specification};
\node (6) [process, below of=1] {Sensor};
\node (7) [process4, below of=6, xshift=-2cm] {$Agent_1$};
\node (8) [process4, below of=6, xshift=+2cm] {$Agent_n$};
\node (21) [process21, above of=1, xshift=-3.5cm,yshift=+3.8cm] {FO Logic};
\node (22) [process21, right of=21,yshift=-2.4cm,xshift=+2.3cm] {Property};
\node (20) [process21, below of=22] {Current State};
\node (23) [process21, above of=22 ] {Constraint};
\node (24) [process21, above of=23, xshift=-1cm ] {Sent Msg};
\node (25) [process21, above of=23, xshift=+1cm ] {Action};
\node (26) [process21, above of=24, xshift=+1cm ] {Received Msg};
\node (27) [process21, above of=26] {Updated State};
\node (28) [process21, above of=27] {Updated Property};
\node (29) [process21, above of=28] {Safety Checking};
\draw [arrow] (21) to [out=270,in=180](22.west);
\draw [arrow] (20) --(22);
\draw [arrow] (22) --(23);
\draw [arrow] (23) --(24);
\draw [arrow] (23) --(25);
\draw [arrow] (21) to [out=60,in=180] (26.west);
\draw [arrow] (24) to [out=140,in=180] (27.west);
\draw [arrow] (25) to [out=30,in=360] (27.east);
\draw [arrow] (26) -- (27);
\draw [arrow] (27) -- (28);
\draw [arrow] (28) -- (29);
\draw [arrow] (6) --(1);
\draw [arrow] (7) --(6);
\draw [arrow] (8) --(6);
\draw [arrow] (1) --(0,0.8);
\draw [arrow] (3) --(-3.2,0.8);
\draw (-5,0.8) rectangle (4.6,10.6);
\draw (-4,10.2) node{DM Node};
\draw [arrow] (29) to [out=180,in=180] node[anchor=east] {Safe }(7.west) ;
\draw [arrow] (29) to [out=0,in=0] node[anchor=west] {Safe }(8.east) ;
\draw [arrow] (29) to [out=0,in=0] node[anchor=west] {Unsafe }(20.east) ;
\end{tikzpicture}
    \caption{ Decision-Making Node (Architecture)}
    \label{fig:DMA}
\end{figure}

Figure \ref{fig:DMA} presents the architecture for the decision-making node in ROS. For a multi-agent system, the decision-making node is expected to generate real-time safe decisions.

Using a multi-agent system consisting of n agents, we demonstrate the similarities and differences between the improved static verification approach and the proposed runtime verification approach.

In terms of the specification language, these system-level specifications are predefined by users in both the static verification approach and the runtime verification approach. However, the agent-level specifications are differently provided in these two approaches. Users and sensors provide the agent-level specifications in the static verification approach and the runtime verification approach, respectively.

The static verification approach and the runtime verification approach share the same implementation of the decision-making generation, but they also contain many modifications to state-space generation and verification. In terms of state-space generation, the entire state space is generated at once via the static verification approach. Specifically, the static method generates every agent decision that could be made to move the system from the initial state to the desired state. The runtime verification approach, on the other hand, only generates one decision at a time, moving from the current state to the following state. In terms of verification, the static verification approach employs probabilistic model checking, using the existing efficient model checkers to verify the properties of the whole state space. The runtime verification approach, on the other hand, merely uses invariant checking while benefiting from intrinsically generated properties during the decision-making generation. As in the static verification approach, each substate is automatically associated with its property. The property checking of the decision-making node can be as simple as checking that all safety specifications are contained in the substate property. The time complexity of this invariant-checking operation is anticipated to be $O(1)$.

Real-time data processing is the key difference between the static verification approach and the runtime verification approach. Real-time data cannot be processed by the static verification method, but they can be processed by the runtime verification approach. Sensors send the decision-making node the real-time beliefs and goals of the agents within the multi-agent system. The decision-making generation module anticipates the ground predicates that the sensor data has been transformed into by the data processing module. The decision-making node generates real-time decisions based on real-time beliefs and goals. Real-time information will presumably be beneficial for decision-making generation for more accuracy.
\subsection{Analysis}
As the first attempt to transition the static verification approach to the runtime verification approach, the proposed runtime verification approach is expected to be a complementary approach to the improved static verification approach. The main core of the runtime verification approach can hugely benefit from the implementation of the static verification approach. The proposed approach can provide safety assurance at runtime, and it does not suffer from the state-space explosion. 

In our setting, each multi-agent system only facilitates one decision-making node, and the decision-making node can only make sequential decisions. Presumably, parallelism could be a bottleneck of the proposed runtime verification approach. 

Furthermore, we point out one important issue in the runtime verification approach. We need to generate a transformation between the high-level decision-making and low-level abilities of autonomous agents to enable communication between the decision-making node. The granularity of the decision-making should be properly designed. If the granularity is too high, the decision-making node has to handle too much data, which will incur unnecessary computational costs and reduce the efficiency of generating decisions. If the granularity is too low, the decision-making node has insufficient control over the autonomous agents. 
 
\section{Conclusion and Future Work}
This paper provides a research preview on how to generate safe autonomous decision-making in ROS. In summary, the proposed runtime verification will be achieved by a two-stage extension of our existing static verification approach. First, we describe how to extend our existing static verification approach to verify multi-goal multi-agent autonomous decision-making. Second, we explain how to transition the improved static verification approach to the proposed runtime verification approach. The main component of the runtime verification approach is a decision-making node. We provided an overview of the architecture of the decision-making node. The decision-making node is expected to generate safe decisions at runtime. We also mentioned that parallelism could be the possible bottleneck of the proposed runtime verification.

We have implemented the initial version of the improved static verification approach, and we validated the initial implementation through two case studies. Our case studies are based on the GOAL source code, provided in the GOAL example project Tower and Coffee of the Eclipse plugin for GOAL. Moreover, we point out that the development of the decision-making node can hugely benefit from the existing implementation of the static verification approach, because they share the same implementation of the decision-making generation. Notably, the static verification approach generates all feasible states from the initial state to the desired state at a time, whereas the runtime verification approach only generates the following state. The static verification approach requires the probabilistic model checking of the whole state space, while the runtime verification approach only requires invariant checking. Therefore, the runtime verification approach is expected to have better performance concerning efficiency than the static verification approach. Initial tests confirm this assumption, yet a thorough analysis will have to validate our hypothesis.

We will fully implement the proposed runtime verification approach in the near future. After that, we will use a multi-agent system consisting of three autonomous mobile robots to validate the improved static verification approach and the runtime verification approach. We also plan to prove the correctness of our implementation of the decision-making node in the future. 
\section*{Acknowledgements}
This research is partially funded by the Research Fund KU Leuven.
\bibliographystyle{eptcs}
\bibliography{generic}
\end{document}